# High Performance Canny Edge Detector using Parallel Patterns for Scalability on Modern Multicore Processors

**Hope Mogale**
North-West University, NW, RSA
Hope.Mogale@ieee.org

## ABSTRACT

Canny Edge Detector (CED) is an edge detection operator commonly used by most Image Feature Extraction (IFE) Algorithms and Image Processing Applications. This operator involves the use of a multi-stage algorithm to detect edges in a wide range of images. Edge detection is at the forefront of image processing and hence, it is crucial to have at an up to scale level. Multicore Processors have emerged as the next solution for tackling compute intensive tasks that have a high demand for computational power. Having significant changes that restructured the microprocessor industry, it is evident that the best way to promote efficiency and improve performance is no longer by increasing the clock speeds on traditional monolithic processors but by adopting and utilizing Processors with Multicore architectures. In this paper we provide a high performance implementation of Canny Edge Detector using parallel patterns for improved performance and Scalability on Multicore Processors. The results show significant improvements in overall performance and this proves that our implementation using parallel patterns does not under utilize resources but scales well for multicore processors.

## 1 INTRODUCTION

Advanced Vector Extensions (AVX) enable Single Instruction Multiple Data (SIMD) technology on Modern Multicore Processors. Moore's Law still continues to provide performance in the multicore era [1]. The evolution of Multicore microprocessors which took over the traditional monolithic processors eradicating the Pentium V dream over the years was due as a result of several limits architects encountered over the years. Several of these limits could be ignored however, the most notable ones that architects could not ignore are the ones discussed by authors in [2] known as the three walls. The first of these walls to be encountered or realized was the power wall as a result of unacceptable growth in power usage with clock rate and also the realization was that above around 130W air cooling is insufficient [3], [2]. Second was the Instruction-level Parallelism (ILP) wall and third was the Memory wall which resulted because processor speeds were highly discrepant to memory speeds. Significant wall of all of the three walls was the power wall. This wall caused so much impact in the microprocessor industry and perhaps was the most compelling to architects to shift processor designs to Multicore Architectures and this successfully led to the establishment of the Multicore processor era even though this also opened the door for Amdahl 's law.

Canny Edge Detector (CED) is a operator used commonly for image feature extraction and also adopted by many image processing algorithms. This operator involves the use of a multi-stage algorithm to detect a wide range of edges in images. CED named after its author [4] nominates a computational approach to edge detection.

Canny Edge Detector operator mainly aims at achieving:

1) Low error rate - Reliable for accurate detection of only existent edges. For low error rates which yields good detection canny edge detector uses Signal-To-Noise (SNR) ratio and its criterion for low error rates [5] on detection is:

$$SNR = A\left|\int_{-T}^{0} f(x)dx\right| \Big/ \left(\sigma\sqrt{\int_{-T}^{T} f^2(x)dx}\right)$$

2) Good localization - The distance between edge pixels detected and real edge pixels have to be minimized. The criterion for good localization [5] is defined as:

$$L = 1 \Big/ \sqrt{\int_{-T}^{T} y^2 \Pr(y)dy}.$$

3) Minimal response - Restrict only one detector response per edge. In other terms the detector should produce multiple maxima. According to Canny [4] the minimal response criterion is defined by:

$$x_{\max}(f) = 2\pi\sqrt{\int_{-T}^{T} f'^2(x)dx \Big/ \int_{-T}^{T} f''^2(x)dx}$$

Because of the aforementioned attributes CED has been widely adopted for image processing applications that involved edge detection. Research has shown that CED also performs better against Laplacian operator which is known to defined by:

$$Laplace\ (f) = \frac{\partial^2 f}{\partial x^2} + \frac{\partial^2 f}{\partial y^2}$$

Edges of an image are important in determining features of digital images has why CED has been applied in many areas of image feature extraction. For enhanced edge detection, CED uses additional algorithms such as Sobel algorithm which are employed on the multi-stages of CED.

We aim provide an efficient parallel implementation of CED that uses parallel patterns for efficient processing and scalability. For us to achieve this we propose a universal parallel computational model we call the Golden Circle of Parallelism (GCP) that we will use to define our structured approach. The GCP Model is composed of three layers and these layers are structured hierarchically as Shell, Kernel and the Core. Figure 1 presents a sketched representation of the GCP model.

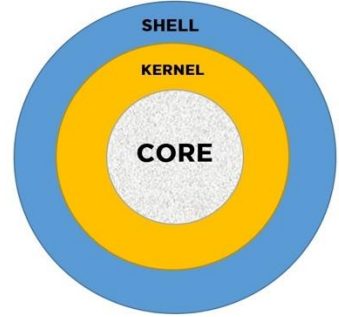

**Figure 1: Golden Circle of Parallelism**

The shell mainly synthesizes any real world problem encountered and pragmatically processes it according to its specific domain. That is to say, the shell synthesizes the problem into an algorithm which relates with its domain. The shell is mainly for denying a problem as an algorithm and identifying opportunities of parallelization for the kernel. When this layer has successfully executed, what is produced is a parallel algorithm that is now ready for the kernel. The kernel plays a critical role on the GCP Model and hence is the most important layer. This layer optimizes the algorithm for the core layer which contains the underlying parallel architectures. This layer is aware of the underlying parallel architecture and hence optimizes for such architecture. For example, if the parallel architecture is a multicore architecture with 8 cores, the kernel layer will optimize and distribute work for each core.

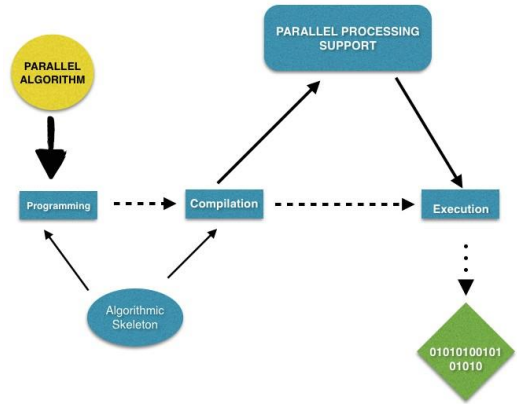

**Figure 2: The kernel Layer**

There are three main phases of the kernel layer, namely programming, compilation, and execution, as can be seen in Figure 2. Each phase has interceptions which highly optimize the phases, and both the programming and compilation phases fully utilize algorithmic skeletons. Also, compilation yields parallel processing support, hence the binaries scheduled for processing are highly parallel and can take full advantage of the underlying parallel architecture. Lastly, we have the core layer which mainly works hand in hand with the underlying parallel architecture of the system. However, this does not mean that the core is tied to any specific object models. The core simply complies with the parallel architecture of the system utilizing the GCP model. In this paper we aim to utilize the GCP Model to enable us to provide a parallel implementation of CED using parallel patterns which can effectively scale on multicore processors and other parallel architectures.

The rest of the paper is organized as follows, Section 2 provides a brief survey of related work, experimental, and computational details. In section 3 we provide results and discussion obtained from the experiments in section 2. Lastly Section 4 provides the conclusion of our work.

## 2 RELATED WORK AND EXPERIMENTAL AND COMPUTATIONAL DETAILS

### 2.1 Selected Related Research Work

There has been substantial amount of research that accounts for CED and its application to image processing operations published by scholars in the past and recently. Also since researchers in [6] evaluated and demonstrated that CED per- forms better than other edge detectors we shall restrict our focus to work that is based on implementation of CED.

Ali et al [7] implemented CED for feature extraction on remote sensing images and recommended CED as an enhancement tool that can be used for troublesome remote sensing images that can be corrupted by point noise. As illustrated by the authors in [8] IFE is a time consuming process and this is even worse when the images to be processed are in large quantities of if the image has high quality. Consequently this is often encountered by image processing applications on the INTERNET because of the high frequency of image data on the INTERNET. Researchers in [9] have identified that the calculation of IFE algorithms constantly increases, and this contributes the most time consuming step in image steganography detection. Researchers in [8] discovered that most IFE methods do not care much about performance and do not take note on the utilization of the highly developed microprocessor architectures. Almost all image processing applications are implemented serially and this leads to poor results in terms of performance even on highly developed microprocessor architectures of the Modern day computer systems. Researchers in [10] surveyed existing shape-based feature extraction. Yang and team recommended that efficient shape features must present essential properties such as identifiability, translation and noise resistance among others. They further outlined that in a simple form a shape descriptor is simply a set of numbers that describe a given shape feature, and in one of the requirements of a shape descriptor they state that the computation of distance between descriptors should be simple; otherwise execution time will present overhead. Since the descriptor operates in serial and is not optimized for multicore architectures this may still be prevalent on application making use of the descriptor.

Kornaros et. al [11] explored several micro architectural alternatives to improve performance for edge detecting algorithms and they proposed reconfigurable multicore prototype which was able to achieve 5x speed up rates. Hao and team in [12] successfully parallelized a Scale Invariant Feature Transform. In their work they state that in order to meet computation demands they optimized and parallelized SIFT to accelerate its performance on Multicore Architecture systems. Furthermore, they indicate that SIMD integrated with Multicore Architectures bring an extra 85% performance increase. Luo et. al [13] successfully implemented CED on NVIDIA Compute Unified Device Architecture (CUDA) and this shows that CED can also be implemented on GPU platforms which is relevant because multicore processors have GPU unit. Zhang et. al [14] presented an improved parallel SIFT implementation which is able to process video images in real-time utilizing multicore processors and the results showed great improved in terms of speedup in comparison to GPU implementation. Cho and team in [15] spearheaded a study that construed the key factors used in the design and evaluation of image processing algorithms on massive parallel platforms. Clemons et. al [16] presented an embedded multicore design named EFFEX with novel functional units and memory architecture support capable of increasing performance on mobile vision applications while lowering power consumption rates. In addition to all these studies we have also identified researchers that have undertaken numerous parallel approaches to implementing CED efficiently. however, all of their undertaken approaches does not use parallel patterns for efficient implementation and computation of CED. Instead their approaches mainly rely on the advantage of parallel hardware for high performance. As an example, Nokano et. al in [17] implemented CED mainly for GPU hardware using CUDA and shows that the same implementation does not scale well for CPUs. The same was for Vassiliadis et. al in [18] when they prosed a parallel implementation of CED with a design that was synthesized for low-end and high-end Xilinx FPGAs, and it achieved a total rate of 240 frames per second for IMpixel images on a Spartan-3E.

Our approach is novel in that we do not rely on specific parallel hardware for performance and scalability but rather on structured parallel patterns which were designed to gain performance and provide scalability on any underlying parallel hardware. We did this because we believe that the ultimate goal of high performance computing is not to obtain performance on certain cores or nodes but to maintain parallelism on any underlying parallel architecture regardless of its hardware capabilities and configuration.

## 2.2 Experimental and computational details

For the experimental setup of our High Performance Canny Edge detector we have chosen a set of images to test performance of our Parallel canny operator. The hardware configuration we have chosen to prepare consists of two common multicore processors of the same architecture listed on table 1. below:

**Table 1: Hardware Configuration for experiment**

| Processor | Vendor | Core Count | Clock Speed |
|---|---|---|---|
| Core i3 | Intel | 2cores, 4 CPUs | 3.4 GHz |
| Core i7 | Intel | 4cores, 8 CPUs | 3.4 GHz |

We have fully parallelized the CED operator to take full advantage of the available resources of any given multicore processor. The main goal is not to have all cores overclocked but rather to have even distribution of work across all cores as seen in figure 3 while aiming for deterministic output.

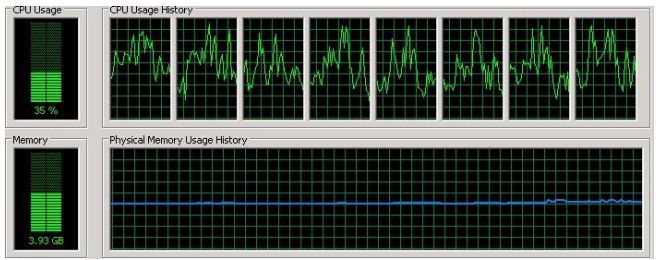

**Figure 3: Even distribution of load on cores**

To achieve deterministic output we have utilized Parallel patterns included with Cilk Plus and we have applied it directly on the Gaussian filter and on Sobel's algorithm which finds the intensity gradient of the image.

To obey Amdahl's law the hysteresis part of the CED algorithm has been left unparallelized and noted to contribute unparalleled work because of the serial elision it carries. That is because of the if statement pattern this forces serial work on the design of the algorithm. CED algorithm was implemented on Microsoft Visual Studio extended with OpenCV and Intel Cilk Plus and was applied on different applications of different disciplines to see how it performs. For all these research areas both the optimized algorithm and the non optimized of CED has been carried out on the images and the results were recorded for each

### 2.2.1 Parallel Canny Edge Detector

The canny edge detector algorithm we have chosen to use is well known and has been widely adopted by many authors in many fields of image processing this operator was first defined by its author in [4] illustrated in figure 4 and has been improved and modified by researchers over time most notably authors in [5].

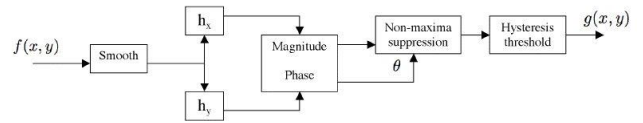

**Figure 4: Canny Edge detector description**

A description of the parallel implementation of the Canny Edge Detector is given below. The formal description can be seen defined in algorithm 1 listing shown below.

---
**Algorithm 1** Canny Edge Detector with Cilk Parallel Patterns
---
1: **procedure** $cilk\_spawn$ **gaussianfilter**(noise $x$)
  ▷ *detect the remaining edges after filtering $x$*
2:   **for** $i = 1 \to k$ **do**
3:     **if** $f_i(x)\%N == 0$ **then**
4:       **return**
5:     **end if**
6:   **end for**
  ▷ *Employ Sobel algorithm*
7:   **for** $i = 1 \to k$ **do**
  ▷ *Parallel loop ($cilk\_for$)*
8:     **if** threshold $p >0$ **then**
9:       $p \leftarrow G_{(x,y)}$
10:      $\theta \leftarrow \arctan\left(G_{(x)} \div G_{(y)}\right)$
11:    **end if**
12:  **end for**
  ▷ *Perform hysterisis*
13: **end procedure**

From the Algorithm listing our parallel CED algorithm can be summarized using the following steps:

**Step 1: Filter out any noise**

We apply parallel patterns seen in figure using Cilk Plus to the Gaussian noise filter. We do this to target deterministic output for our operator

**Step 2: Obtain Gradient intensity of the image**

To achieve this step, we employ the Sobel Algorithm with parallel modifications on the computation. Firstly, we apply parallel patterns to obtain determinism on a nominal pair of convolution masks usually denoted by (*Gx, Gy*). Then lastly the gradient strength and its direction is realized through parallel computation.

**Step 3: Apply the Non-Maximum Suppression Filter**

We use the non maximum suppression remove pixels that are not part of the edge. This acts like a low pass filter for unwanted pixels that are not part of the edges.

**Step 4: Perform Hysteresis**

The Hysteresis is used to detect if the pixel gradient is higher, lower or between the upper threshold. If higher then it is accepted, if below then it is rejected and if between then it will be accepted only if it is connected to a pixel that is above the upper threshold. The Hysteresis is usually performed as the last step of the CED algorithm as seen in figure 5.

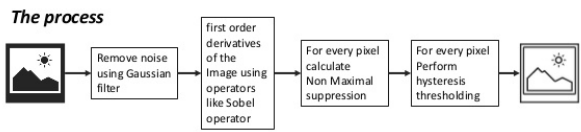

**Figure 5: Canny Edge Detector Process**

From the four steps described above we note that step 4 will be the main cause of forced serial computation on most modern multicore architecture technology. To counter this for throughput we adjust that an asymmetric multicore approach be utilized especially on the serial *(1-f)* part of of the equation illustrated on the corollary of the original fundamental principle of computation by Amdahl.

$$Speedup_{asymmetric}(f, n, r) = \frac{1}{\frac{1-f}{perf(r)} + \frac{f}{perf(r)+n-r}}$$

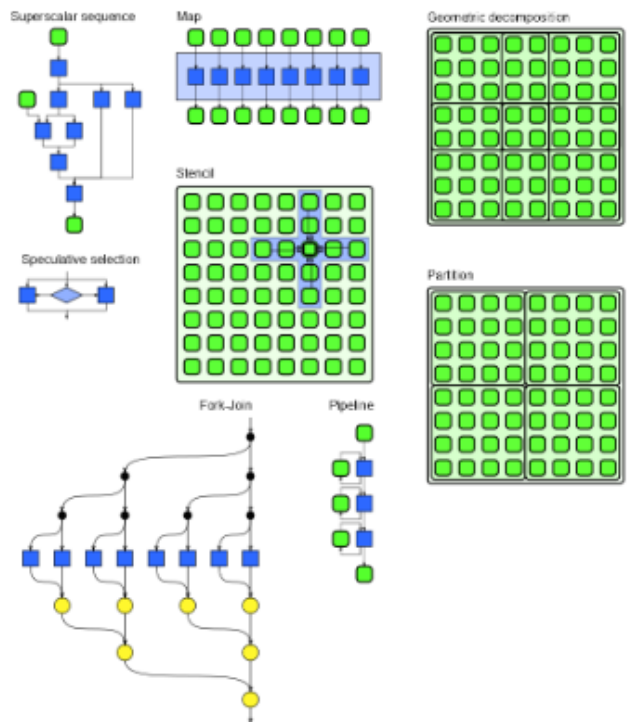

**Figure 6: Parallel Patterns for Deterministic Computation [2]**

### 2.2.2 Parallel Programming Models

For the parallel implementation of CED we have chosen Intel Cilk Plus which provides composable parallel patterns that guarantee determinism. Cilk Plus is a multithreaded language that uses a work stealing scheduler that has the ability to distribute work loads evenly on the cores of the multicore processor. Cilk is highly algorithmic [2] hence, it provides algorithmic skeletons that renders the Cilk runtime to take care of details such as load balancing, resource communication, and etc. Cilk plus features a set of parallel patterns highly suitable for numerical computation. There are many parallel patterns available in Cilk and a few can be seen in figure 6. It is also important to note that Cilk only recommends what can run in parallel. Thread-parallelism in Cilk Plus is expressed with the notion of a strand.

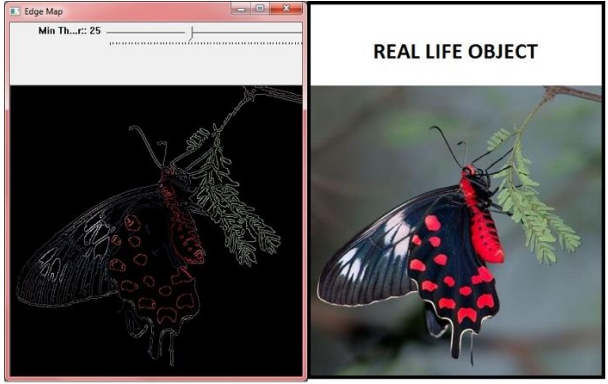

**Figure 7: Parallel CED Algorithm application run**

Our Parallel Canny Edge detector shows positive results as can be seen in figure 7. From the results above it can be be seen that our parallel CED algorithm can play an important role in Computer Vision Applications. In the follow up section we discuss selected obtained results from our experimental setup.

## 3   RESULTS AND DISCUSSION

We now discuss the obtained results from our experimental setup. We discuss both the parallel (optimal) and non-parallel (suboptimal) implementation of the Canny Edge Detector in the context of our CPU types:

### 3.1.  CPU Sampling and Total Usage

1) **Suboptimal implementation:** The sampling method used collects profiling data for every 10,000,000 processor cycles and this is very useful for detecting performance issues. For the suboptimal(non-parallel) or non-optimal implementation of CED algorithm the profiler collected about 8,992 samples. A graphical of representation of CPU usage over wall clock time in seconds is shown in figure 8. From this graph a low CPU usage can be seen which is not ideal for application performance and user experience. The low CPU usage is in response to the total sample count.

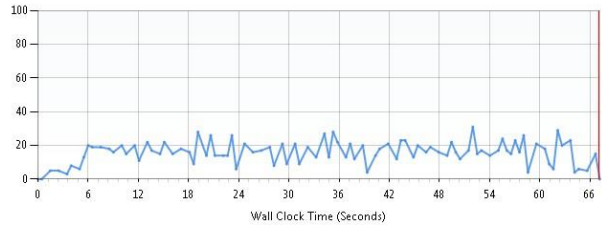

**Figure 8:  Suboptimal CPU Usage Over Wall Clock Time**

2) Optimal implementation:  For the optimal implementation of the CED algorithm the profiler collected about 34,884 samples. In Figure 9 the graphical representation of CPU usage over wall clock time in seconds can be seen for the fully Optimized CED algorithm. This graph shows better CPU usage which is ideal for application performance and user experience. The efficient CPU usage is also in response to the total sample count.

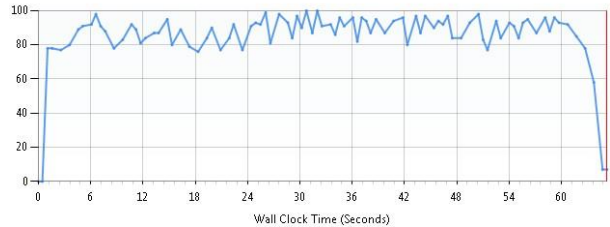

**Figure 9:  Optimal CPU Usage Over  Wall Clock Time**

From the observed results it  can  be  seen that the  optimal CED out  performs  the  suboptimal implementation  of  the  CED algorithm. The above results provides only information for total CPU Usage in percentage. To be precise the total usage per core must be observed. Next we observe total usage per core for 4 core CPU and 8 core CPU to test the developed CED for scalability.

**3.2 Total CPU Usage Per Core**

1) **Suboptimal implementation (4  Cores)**:  To  be  precise if the CED algorithm is stable and durable in terms of performance total usage per core must be observed. In figure 9 this can be seen precisely, the total CPU usage per core for the suboptimal implementation. From the figure it can be seen that the utilization is uneven hence, declaring that some cores may be idle while others are working. This is not ideal for overall application performance on a Multicore Architecture system.

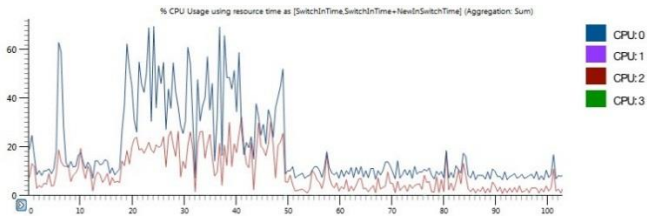
**Figure 9: Suboptimal CPU Usage Per Core(4 CPUs)**

2) **Suboptimal implementation (8 CPUs)**: In figure 10 we also show the suboptimal implementation of CED with no parallelism. From the figure it can be seen that this implementation is not ideal for a Multicore system with more CPUs.

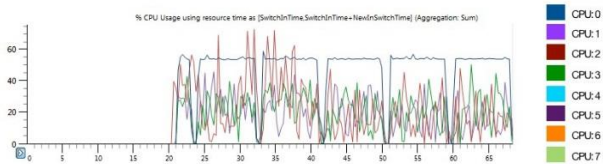
**Figure 10: Suboptimal CPU Usage Per Core(8 CPUs)**

3) **Optimal implementation (4 CPUs)**: The total usage per core for the Optimal implementation of CED on 4 CPU multicore processor is shown figure 11 and it can be see that there is an ideal usage per core. From the figure, we can see that utilization per core is balanced and evenly distributed among all the cores. This is mainly because of the work stealing scheduler provided by Cilk Plus runtime. This even distribution is ideal for Multicore Architecture based systems and will enhance overall application performance and user experience.

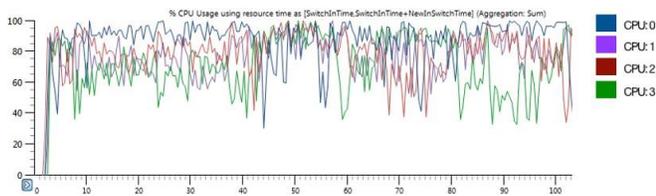
**Figure 11: Optimal CPU Usage Per Core(4 CPUs)**

4) **Optimal implementation (8 CPUs):** To test if our CED implementation is scalable we have implemented it on the 8 CPU multicore processor and the results can be seen in figure 12. The results show ideal usage that fully maximizes and utilizes available CPU resources. The results seen in figure 4.6 serve as proof that our parallel implementation of CED is fully scalable for multicore processors. The uneven peaks seen in figure 12 and other figures demonstrate the breath in effect of the Cores since they are using too much power when utilized all at once.

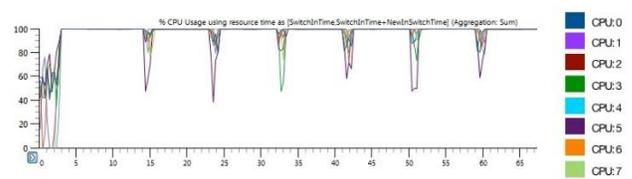
**Figure 12 : Optimal CPU Usage Per Core(8 CPUs)**

From our obtained results we have demonstrated that as proof of concept our parallel CED algorithm has high performance output.

## 4 CONCLUSIONS

In this paper we have successfully utilized parallel patterns to implement and demonstrate a Scalable High Performance Parallel implementation of Canny Edge Detector. Through a defined structural approach like GCP, we have proved that the parallel implementation of CED is highly optimal for Multicore Processors. The results showed improved CPU usage over wall clock time, and efficient CPU usage per core. The results were obtained for both Multicore processor systems with 4 and 8 CPUs. The results showed effective performance of parallel implementation of CED on both Multicore processor systems hence proving that the parallel implementation of CED is scalable. The only challenge encountered is with regards to power usage which can be drastically high when all cores are utilized. To counter this, we recommend that an asymmetric approach be used for application, which will require longer times of full usage of all cores such as high resolution image and video rendering. In future we aim to further extend our implementation of the parallel CED on many core systems with 32-64 CPUs to test for robustness of the parallel CED.

## ACKNOWLEDGMENTS

We would like to thank everyone who took part and made effort to provide all the support and infrastructure to make this work possible.